\documentclass[prl,twocolumn,showpacs]{revtex4-1}
\usepackage{amsmath,amssymb}
\usepackage{graphicx,color}
\newcommand{\bra}[1]{ \langle #1 |}
\newcommand{\ket}[1]{| #1 \rangle}
\newcommand{\bk}[2]{\left \langle #1 | #2 \right \rangle}
\newcommand{\aver}[1]{\left \langle #1 \right \rangle}
\renewcommand{\Im}{\textrm{Im}}

\newcommand{\Heff}{\mathcal{H}_{\mathrm{eff}}}

\begin{document}
\title{Probing eigenfunction nonorthogonality by parametric shifts of resonance widths}
\author{D. V. Savin} \author{J.-B. De Vaulx}
 \affiliation{Department of Mathematical Sciences, Brunel University, Uxbridge, UB8 3PH, United Kingdom}
\published{in \emph{Acta Phys. Pol. A \textbf{124}, 1074 (2013)}\,}
\begin{abstract}
Recently, it has been shown that the change of resonance widths in an open system under a perturbation of its interior is a sensitive indicator of the nonorthogonality of resonance states. We apply this measure to quantify parametric motion of the resonances. In particular, a strong redistribution of the widths is linked with the maximal degree of nonorthogonality. Then for weakly open chaotic systems we discuss the effect of spectral rigidity on the statistical properties of the parametric width shifts, and derive the distribution of the latter in a picket-fence model with equidistant spectrum.
\end{abstract}
\pacs{03.65.Nk, 05.45.Mt, 05.60.Gg }        
\maketitle
\setlength{\unitlength}{1cm}
\begin{picture}(0,0)(-1.55,-0.65)
 \put(0,0){\parbox{0.8\textwidth}{\footnotesize Proceedings of 6th Workshop on Quantum Chaos and Localisation Phenomena, Warsaw, May 24--26, 2013}}
\end{picture}
\vspace*{-5ex}

\subsection{1. Introduction}

Addressing resonance phenomena in open systems, one usually adopts  the scattering approach \cite{Mahaux} based on the so-called effective non-Hermitian Hamiltonian \cite{verb85,soko89,fyod97,rott09,fyod11ox}
\begin{equation}\label{Heff}
 \Heff = H - \frac{i}{2} AA^\dag\,.
\end{equation}
Its Hermitian part $H$ gives rise to $N$ energy levels of the closed system. The anti-Hermitian part is responsible for their coupling to $M$ open scattering channels, with $A$ being an $N{\times}M$ matrix of decay amplitudes. The scattering  resonances are then given by the complex eigenvalues $\mathcal{E}_n=E_n-\frac{i}{2}\Gamma_n$ of $\Heff$, with energies $E_n$ and widths $\Gamma_n>0$. Since $\Heff$ is non-Hermitian, the corresponding right and left eigenfunctions, $\Heff\ket{R_n} = \mathcal{E}_n\ket{R_n}$ and $\bra{L_n}\Heff = \mathcal{E}_n\bra{L_n}$, are no longer orthonormal but rather form a biorthogonal system, satisfying the relations $\bk{L_n}{R_m}=\delta_{nm}$ and $\sum_{n=1}^N\ket{R_n}\bra{L_n}=1$.

Nonorthogonality of such resonance states can be generally quantified by the matrix $U_{nm}=\bk{L_n}{L_m}$ that was first introduced by Bell and Steinberger \cite{bell66} in nuclear physics (see also \cite{soko89,soko9497}). This matrix differs from the unit matrix; it influences the decay laws of open systems \cite{savi97} and appears in other physical applications \cite{chal98}. For example, the \emph{diagonal} element $U_{nn}$ is known in optics as the Petermann factor of a lasing mode \cite{pete79,schom00a}. The other (related to $U_{nn}$) characteristics include the phase rigidity \cite{kim05,bulg06b} and the mode complexness \cite{savi06b,poli09b} in open microwave cavities. Nonorthogonal mode patterns also emerge in optical microstructures \cite{wier0811} as well as in reverberant dissipative bodies \cite{lobk00a} and elastic plates  \cite{xeri09}.

Non-Hermitian operators are generally known to exhibit extreme sensitivity to perturbations \cite{tref97}. For open quantum systems, an important connection has been very recently recognized in \cite{fyod12b}, establishing a parametric shift of resonance widths as a sensitive measure of nonorthogonality. Namely, one considers the parametric motion of resonance states, described Eq.~(\ref{Heff}), under a perturbation of the internal region. This can be modelled by
\begin{equation}
 \Heff \rightarrow \Heff'=\Heff +\alpha V\,,
\end{equation}
where $V$ is a Hermitian $N\times N$ matrix and $\alpha$ is a real parameter controlling the perturbation strength. The shift $\delta\mathcal{E}_n$ of the $n$th resonance can then be found by applying a perturbation theory routine with necessary modifications induced by biorthogonality \cite{fyod12b,Kato}. To the first order in $\alpha$ this readily yields the resonance shift as
\begin{equation}\label{shift}
 \delta\mathcal{E}_n \equiv \mathcal{E}_n'-\mathcal{E}_n = \alpha\bra{L_n}V\ket{R_n}\,,
\end{equation}
generalizing the standard result to the non-Hermitian case. It is the eigenfunction nonorthogonality that causes a nonzero value of the imaginary part of $\delta\mathcal{E}_n$. This fact can be clearly seen from the following representation for the parametric shift of the resonance width \cite{fyod12b}:
\begin{equation}\label{dgam}
 \delta\Gamma_n \equiv-2\Im(\delta\mathcal{E}_n) = i\alpha\sum_{m}(U_{nm}V_{mn}-V_{nm}U_{mn})\,,
\end{equation}
where $V_{nm}=\bra{R_n}V\ket{R_m}=V_{mn}^{*}$. Since only the terms with $m\neq n$ contribute to the sum above, the width shift (\ref{dgam}) is solely induced by the \emph{off-diagonal} elements $U_{nm}$ of the nonorthogonality matrix, thus vanishing only if the resonance states were orthogonal.

In this work, we study the general features of this new nonorthogonality measure in the context of parametric motion of two interfering resonances. Then we discuss statistical properties of the width shifts in weakly open chaotic systems with or without spectral fluctuations.

\subsection{2. Unstable two-level system}

We will consider the case of preserved time-reversal symmetry, when both $H$ and $V$ are real symmetric matrices and $A$ is also real. The system in question is generally described by the following effective Hamiltonian:
\begin{equation}\label{Heff2}
 \Heff = \frac{1}{2} \left(\begin{array}{cc}  \Delta-i\gamma_1 & -i\sqrt{\gamma_1\gamma_2}\cos\theta
         \\ -i\sqrt{\gamma_1\gamma_2}\cos\theta  & -\Delta-i\gamma_2 \end{array}\right)
        + \alpha \left(\begin{array}{cc} d  & v \\ v & -d \end{array}\right).
\end{equation}
Here, $\Delta$ stands for the energy separation of two parental levels. Following \cite{soko89}, we have  parameterized the coupling term $-\frac{i}{2}(AA^\dag)$ in terms of the scalar products of two $M$-dimensional vectors of decay amplitudes, $\{A^c_{1,2}\}$, with the angle $\theta$ between them and $\gamma_{1,2}=\sum_{c=1}^M(A_{1,2}^c)^2$. We have also chosen $V$ to be traceless, thus eliminating the trivial total energy shift. Since $V$ is Hermitian, the term $\alpha V$ does not change the total system openness. Altogether, this implies the following sum rules (at any real $\alpha$)
\begin{equation}\label{sum}
 \begin{array}{l}
  E_1 + E_2 = 0,\\[1ex]
  \Gamma_1+\Gamma_2=\mathrm{const} = \gamma_1+\gamma_2
 \end{array}
\end{equation}
for the energies and widths. A formal diagonalization of (\ref{Heff2}) gives the complex resonances explicitly,
\begin{equation}\label{E_12}
 \mathcal{E}_{1,2} = -\frac{i}{4}(\gamma_1+\gamma_2)\pm \frac{1}{2}\sqrt{\epsilon^2-\nu^2}\,,
\end{equation}
with $\epsilon=\Delta+2\alpha d-\frac{i}{2}(\gamma_1-\gamma_2)$ and $\nu=\sqrt{\gamma_1\gamma_2}\cos\theta+2i\alpha v$. When system parameters change, the energies and widths exhibit crossings and anticrossings, see Fig.~1, which were studied in various physical situations \cite{soko9497,magu99,glue02,voly03,hern11,bitt12}.

\begin{figure}[t]
 \includegraphics[width=0.5\textwidth]{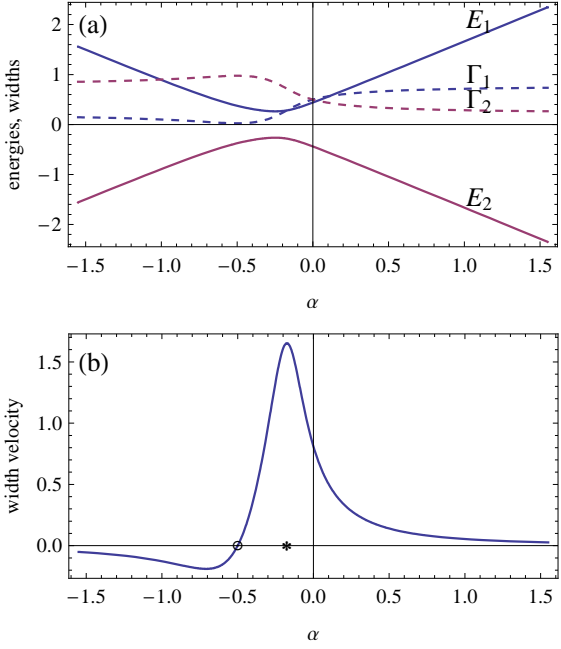}
 \caption{(a) Dynamics of two resonances under the perturbation of the internal region, Eqs.(\ref{Heff2})--(\ref{E_12}), preserving the system openness ($\Gamma_1+\Gamma_2=\mathrm{const}$) and total energy ($E_1+E_2=0$). The energies (solid lines) and widths (dashed lines) are shown as a function of the perturbation strength $\alpha$. The system parameters are: $\Delta=d=1$, $v=0.75$, $\gamma_1=\gamma_2=0.5$ and $\theta=\frac{\pi}{10}$. (b) The corresponding parametric width velocity, Eq.~(\ref{dgam2}). It attains its maximum (or zero) at the value $\alpha_\ast$ (or $\alpha_\circ$) that is indicated on the abscissa with the sign $\ast$ (or $\circ$).}
\end{figure}

To make the connection between such parametric motion and the properties of resonance states, we represent the corresponding right eigenvectors as follows \cite{savi06b} \vspace*{-0.1ex}
\begin{equation}\label{basis}
 \ket{R_1}= \mathcal{N} \left(\begin{array}{c} 1 \\ -if\end{array}\right)\,,
 \qquad
 \ket{R_2} = \mathcal{N} \left(\begin{array}{c} if\\ 1\end{array}\right)\,,
\end{equation}
the left eigenvectors being just the transpose of (\ref{basis}). Here, $\mathcal{N}^2=1/(1-f^2)$ is the normalization constant and we have introduced the complex parameter $f$,
\begin{equation}\label{f}
 f=\nu/(\epsilon+\sqrt{\epsilon^2-\nu^2})\,,
\end{equation}
describing the mixing of the resonance states. In such a parametrization, the Bell-Steinberger matrix reads
\begin{equation}\label{U}
 U = |\mathcal{N}|^2
 \left(\begin{array}{cc}
  1+|f|^2 & -2i\mathrm{Re}f \\ 2i\mathrm{Re}f  &  1+|f|^2
 \end{array}\right)\,.
\end{equation}
Clearly, the nonorthogonality is due to nonzero $\mathrm{Re}f$.

\begin{figure}[t]
 \includegraphics[width=0.5\textwidth]{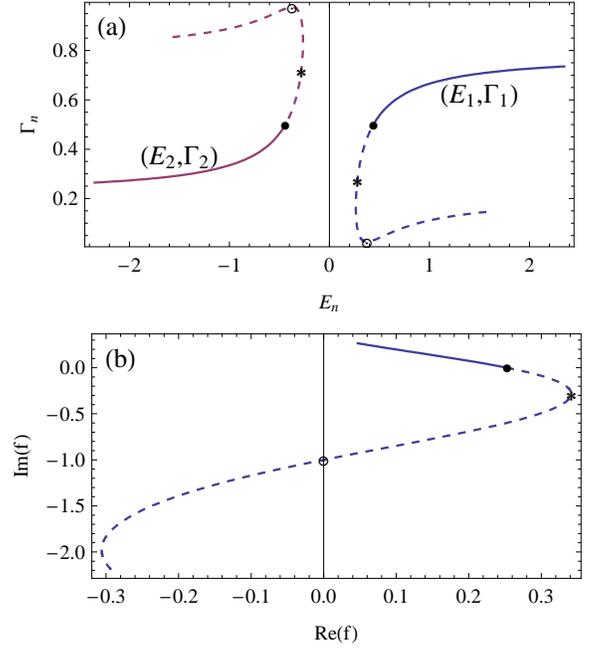}
 \caption{The Argand diagrams for (a) the complex energies $\mathcal{E}_n=E_n-\frac{i}{2}\Gamma_n$; and (b) the mixing parameter $f$. The system parameters are the same as in Fig.~1. The solid (dashed) lines correspond to the positive (negative) values of the perturbation parameter $\alpha$. Dots ($\bullet$) show the initial values ($\alpha=0$). A strong redistribution of the widths is clearly seen at $\alpha=\alpha_\ast$, when the nonorthogonality reaches its maximum (indicated by $\ast$). The corresponding values at $\alpha=\alpha_\circ$, when two resonance states become orthogonal, are shown by $\circ$.}
\end{figure}

It is now instructive to consider the rates, $\dot\Gamma_n\equiv\frac{\delta\Gamma_n}{\delta\alpha}$, at which the widths change with $\alpha\to\alpha+\delta\alpha$ ($\delta\alpha\ll1$). Treating the term $\delta\alpha V$ as a perturbation to (\ref{Heff2}), one can readily find $\dot\Gamma_n$ from (\ref{dgam}) and (\ref{U}) in the explicit form,
\begin{equation}\label{dgam2}
 \dot\Gamma_1 = 4\mathrm{Re}f\frac{v(1-|f|^2)-2d\,\mathrm{Im}f}{(1+|f|^2)^2-4\mathrm{Re}f^2} = -\dot\Gamma_2\,.
\end{equation}
Such a parametric width `velocity' vanishes at $\mathrm{Re}f=0$, when the states are orthogonal. It develops a maximum corresponding to that of $\mathrm{Re}f$, with its height being controlled by $\mathrm{Im}f$. In the vicinity of such a point, when the nonorthogonality is at maximum, a strong redistribution of the widths takes place. All these features are clearly seen on Figs.~1 and 2. Note that this width redistribution has a different nature from that considered in Refs.~\cite{soko89,rott09}, where it was caused by varying the strength of coupling to the continuum. Here, the coupling to the continuum is kept fixed, the width redistribution being induced by interior perturbations due to the increased mixing and nonorthogonality of the resonance states. It is worth mentioning that the point of the maximal degree of nonorthogonality does not generally coincide with that of the minimal distance of the eigenvalues in the complex plane. (See also \cite{brod13} for the general geometric approach to parametric sensitivity in non-Hermitian systems.)

Expression (\ref{dgam2}) together with the corresponding result for the parametric energy velocities, $\dot{E}_n\equiv\mathrm{Re}(\delta\mathcal{E}_n)/\delta\alpha$,
\begin{equation}\label{dEner2}
 \dot{E}_1 = \frac{(1+|f|^2)(d(1-|f|^2)+2v\,\mathrm{Im}f)}{(1+|f|^2)^2-4\mathrm{Re}f^2} = -\dot{E}_2\,,
\end{equation}
provide a direct access to the mixing parameter $f$ from resonance spectra. We stress that $U_{11}$ appears as the proportionality coefficient in (\ref{dEner2}), whereas $U_{12}$ does in (\ref{dgam2}). This gives a promising way of probing spatial characteristics in open systems by purely spectroscopic tools, with various  spectral data being readily available \cite{glue02,bitt12,kuhl08}.

The above description generally holds everywhere except at exceptional points \cite{Kato,rott09,brod13,heis12}. These are the branching points of (\ref{E_12}) corresponding to $\epsilon=\pm\nu$. At such points, $f=\pm1$, which implies coalescence and self-orthogonality of the eigenstates \cite{heis12}. In our model, the proximity to the exceptional points is controlled by $v\neq0$ (for the real parameters). Thus, our results (\ref{dgam2}) and (\ref{dEner2}) also provide analytical tools to study such proximity effects in resonance dynamics, see \cite{eleu13b} for the related study.

\subsection{3. Weakly open chaotic systems}

We proceed with the general case of $N$-level systems in the regime of weak coupling to the continuum.   In this case, the non-Hermitian part of $\Heff$ can be treated as a perturbation to the Hermitian part $H=\sum_nE_n\ket{n}\bra{n}$. To the leading order in the coupling, the resonance widths are given by $\Gamma_n = \sum_{c=1}^M |A_n^c|^2$, whereas the parametric width velocities read \cite{fyod12b}
\begin{equation}\label{dgamN}
  \dot\Gamma_n = \sum_{m\neq n}\frac{\bra{m}G_n\ket{m}}{E_n-E_m}\,,
\end{equation}
where $E_n$ are the energy levels of the closed system and $G_n$ denotes the following Hermitian operator:
\begin{equation}\label{G_n}
  G_n=AA^\dag\ket{n}\bra{n}V+V\ket{n}\bra{n}AA^\dag\,.
\end{equation}

Considering chaotic systems, one usually deals with statistical modeling in the limit $N\gg1$. In the present case of preserved time-reversal symmetry, the coupling amplitudes are chosen as real Gaussian random variables with zero mean and variance $\aver{A_n^aA_m^b}=\overline{\Gamma}\delta_{nm}\delta^{ab}$ \cite{soko89}. This yields the well-known Porter-Thomas distribution,
\begin{equation}\label{PTdis}
 P_M(\kappa) = \frac{1}{ 2^{M/2}\Gamma(M/2)}\kappa^{M/2-1}e^{-\kappa/2}\,,
\end{equation}
for the widths $\kappa_n=\Gamma_n/\overline{\Gamma}$ in units of the mean partial width $\overline{\Gamma}$. With the assumption of Gaussian distributed wavefunctions, it can be shown \cite{fyod12b} that the following representation holds for the rescaled width velocities:
\begin{equation}\label{y_n}
 y_n = \frac{\dot{\Gamma}_n}{\overline{\Gamma}\sqrt{\mathrm{Tr}(V^2)}} = \frac{\sqrt{\kappa_n}}{\pi} \Delta\sum_{m\neq n}\frac{z_m v_m}{E_n-E_m}\,.
\end{equation}
Here, $\Delta$ is the mean level spacing (near the $n$th level) and the quantities $z_m$ and $v_m$ are real normal variables.

The statistical properties of the width velocities $y_n$ can be characterized by the probability distribution function $\mathcal{P}_M(y)=\Delta\aver{\sum_{n=1}^N\delta(E_n)\delta(y-y_n)}$. In the weak coupling regime, spectral and spatial fluctuations become statistically independent that allows one to perform the averaging over $\{E_n, \kappa_n\}$ and $\{z_m,v_m\}$ separately. Making use of the convolution theorem, the final expression for the distribution can be cast as follows \cite{fyod12b}
\begin{equation}\label{shiftdist}
 \mathcal{P}_M(y) = \int_0^{\infty} \frac{d\kappa}{\sqrt{\kappa}}\,
                    P_M(\kappa)\, \phi\left( \frac{y}{\sqrt{\kappa}}\right)\,,
\end{equation}
where the function $\phi(y)$ depends only on the spectral properties of $\{E_m\}$ (around $E_n=0$) and is defined as
\begin{equation}\label{phi}
 \phi(y) = \int_{-\infty}^{\infty}\frac{d\omega}{2\pi} e^{i\omega y}
 \aver{\prod_{m\neq n}\frac{|E_m|}{\sqrt{E_m^2+\omega^2\Delta^2/\pi^2}}}\,.
\end{equation}

Conventionally, the energy levels in chaotic systems with time-reversal symmetry are induced by the so-called GOE-distributed random Hamiltonian $H$ \cite{verb85,soko89}. For such a model, the exact form of $\phi$ was also derived in \cite{fyod12b},
\begin{equation}\label{phigoe}
\phi^{\textrm{(goe)}}(y) = \frac{4 + y^2}{6(1+y^2)^{5/2}}\,.
\end{equation}
When substituted into (\ref{shiftdist}), it leads to the distribution of the width velocities in the GOE case, $\mathcal{P}^{\textrm{(goe)}}_M(y)$. The latter has a power law decay, $\mathcal{P}^{\textrm{(goe)}}_M(y) \propto |y|^{-3}$, which can be linked with the linear level repulsion.

To study the influence of level fluctuations onto statistics of the width velocities, it is instructive to consider the picket-fence model \cite{poli09b}. In this model the energy levels are equally spaced, $E_n-E_{n\pm k}=\pm k\Delta$, implying complete spectral rigidity. The variance of $y$ can be easily computed from (\ref{y_n}) by taking into account the normal character of $z$ and $v$:
$\mathrm{var}(y) = \frac{\aver{\kappa}}{\pi^2}\sum_{k\neq0}k^{-2} = \frac{M}{3}$.
In contrast to the GOE case where all the (even) moments diverge, the finite variance in the picket-fence model implies much faster decay of the corresponding distribution $\mathcal{P}^{\textrm{(pf)}}_M(y)$. To find the latter explicitly, we first note that the product featuring in (\ref{phi}) can now be computed as follows
$ \prod_{k=1}^{\infty}[1+\omega^2/(\pi k)^2]^{-1} = |\omega|/\sinh|\omega|$. Taking the Fourier transform, we finally arrive at
\begin{equation}\label{phipf}
 \phi^{\textrm{(pf)}}(y) =  \frac{\pi}{2[1+ \cosh(\pi y)]}\,.
\end{equation}
Expressions (\ref{shiftdist}) and (\ref{phipf}) give the distribution $\mathcal{P}^{\textrm{(pf)}}_M(y)$ of the width velocities in the picket-fence model. As $M$ grows, this distribution gets broader, approaching the limit $\mathcal{P}^{\textrm{(pf)}}_{M\gg1}(y)=\frac{1}{\sqrt{M}}\phi^{\textrm{(pf)}}(\frac{y}{\sqrt{M}})$ at $M\gg1$. Such a behavior is illustrated on Fig.~3 at several values of $M$.

\begin{figure}[t]
\includegraphics[width=0.475\textwidth]{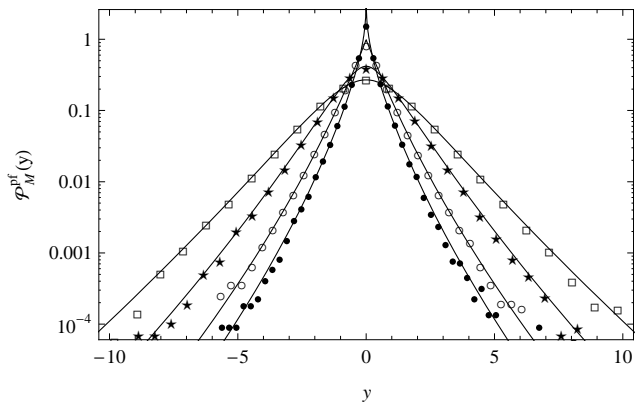}\\
\caption{Distributions of the parametric width velocities for weakly open chaotic systems with the equidistant spectrum at $M=1\,(\bullet), 2\,(\circ), 5\,(\star)$ and $10\,(\Box)$ open channels. The solid lines show the analytical result, Eqs.~(\ref{shiftdist}) and (\ref{phipf}). The symbols stand for numerics with $2000$ realizations of $250{\times}250$ random matrices (only 25 levels around $E=0$ were kept).}
\end{figure}

Comparing the two models, we see that the main impact of spectral fluctuations is on the distribution tails, which become exponentially suppressed in systems with a completely rigid spectrum. By virtue  of (\ref{dgam}), this directly applies to the off-diagonal elements $U_{nm}$ of the nonorthogonality matrix. These results complement similar findings \cite{poli09b} on the diagonal elements $U_{nn}$, thus providing a complete description of statistics of nonorthogonality in weakly open chaotic systems.

\subsection{4. Conclusions}

In summary, we have studied the nonorthogonality of resonance states in open quantum systems by means of parametric dynamics. For the two-level system, we have given the complete analytic solution and, in particular, linked a strong redistribution of the widths with the maximal degree of nonorthogonality. For weakly open chaotic systems, we have found that enhancing spectral rigidity leads to the suppression of nonorthogonality effects.

One of us (DVS) is grateful to I. Rotter and J. Wiersig for helpful discussions and for bringing his attention to Refs.~\cite{eleu13b} and \cite{wier0811}, respectively.

%

\end{document}